# Analysis of quantum tomography protocol efficiency for triphoton polarization states[1]

Yu. I. Bogdanov[2a,b,c], Yu. A. Kuznetsov [c], G. V. Avosopyants[a,c,d], K. G. Katamadze[a,b,d],

L. V. Belinsky[a,c], N. A. Borshchevskaya[d]

[a]Institute of Physics and Technology, Russian Academy of Sciences, 117218, Moscow, Russia.
[b]National Research Nuclear University "MEPHI", 115409, Moscow, Russia.
[c]National Research University of Electronic Technology MIET, 124498, Moscow, Russia.
[d]M. V. Lomonosov Moscow State University, 119991, Moscow, Russia.

**ABSTRACT**

Reliable generation and measurement of triphoton states has yet to be achieved in laboratory. We give an overview of the problems in generating and measuring triphoton quantum states and analyze several protocols of quantum measurements, which allow for high precision of reconstruction when sizes of available statistical data samples are limited. The tomography procedure under investigation is based on root approach to state estimation. In particular, we use the generalized Fisher information matrix to assess the accuracy of the quantum state parameters measurement. We use tomographic protocols, based on the symmetry of the Platonic solids. We demonstrate the capability to reconstruct triphoton quantum states with precision close to the maximum achievable value allowed by quantum mechanics.

## 1. INTRODUCTION

The polarization state of a photon provides convenient means of encoding and broadcasting quantum information. Modern experimental setups allow to prepare, transform and measure polarization states of single photons and correlated pairs (biphotons) with accuracy exceeding 99%.[1–5] However, manipulations with large numbers of photons present significant difficulties. At the moment the accuracy of tripartite entangeled state preparing is less than 86%.[6] This is due to low generation rate of such states – usually less than 1 Hz**.** Here we consider key methods of triphoton preparation and analyze different protocols of the original polarization state reconstruction.

## 2. TRIPHOTON GENERATION METHODS

The basic way to obtain triphoton states is to use the third order spontaneous parametric down conversion (TOSPDC) effect in the medium with third order nonlinearity $\chi^{(3)}$ when one photon of a high intensity pump laser may divide into three photons. However, generation of photons in nonlinear crystals has very low efficiency compared to biphoton generation by second order spontaneous parametric down conversion[7]. According to calculations the generation efficiency

---

of triphotons is of 9 orders lower than generation of biphotons[8] under the same conditions and the generation rate of triphotons is lower than 0.01 Hz.

The alternative media for triphoton generation are optical waveguides with special dispersion (Fig. 1-a).
In that case the theoretical efficiency of generation may reach a few Hz, yet there exists experimental evidence only for the third harmonic generation (THG) there, which corresponds to the more efficient reverse process [9–11]. One may select the appropriate conditions for TOSPDC and THG with the help of microcavities which have very high Q-factor and therefore provide efficient interaction between pump and signal waves[12].
One more promising way of TOSPDC generation is by means of integrated waveguides combining positive properties of both bulk media (high cubic susceptibility) and fibers (the existence of guiding modes)[13].
There are also proposals for triphoton cascade generation in cold atoms[14] and also in quantum dots (the experimental generation frequency reached 1 Hz but the polarization state was not measured[15])
At present the sources of triphoton polarization states have only been successfully realized by means of SPE of second order.

During SPE one can obtain biphoton, four photon and even six photon entanglement, as well as that of higher order depending on the pump intensity.

For tomographic measurements we shall select only the case when all 4 photons are detected. The polarization states of the three photons are measured by the system of phase plates and detectors and the fourth one is the trigger photon. It is used only for selecting the events of simultaneous registration of all four photons. The entangled states presented on Fig.1b were measured with the generation frequency of 7 to 500 mHz[16].

The other way to make triphoton states is based on interference of a pair of spontaneously emitted photons with a quasi-photon state. In this case the extra photon can be appended to one of the channels before or after SPE. That notably increases the efficiency of photon emission.

The maximum generation frequency reached in the first case – 25 Hz, and in the second case – 1.45 Hz[16–22].
The final scheme that we shall consider for triphoton generation is by means of a cascaded SPE using a set of two crystals. The photon generated in the first crystal acts as a pump for the SPE in the second crystal - Fig.1e[6,23,24].
The frequency of triphoton generation do not exceed few tenths of Hz.

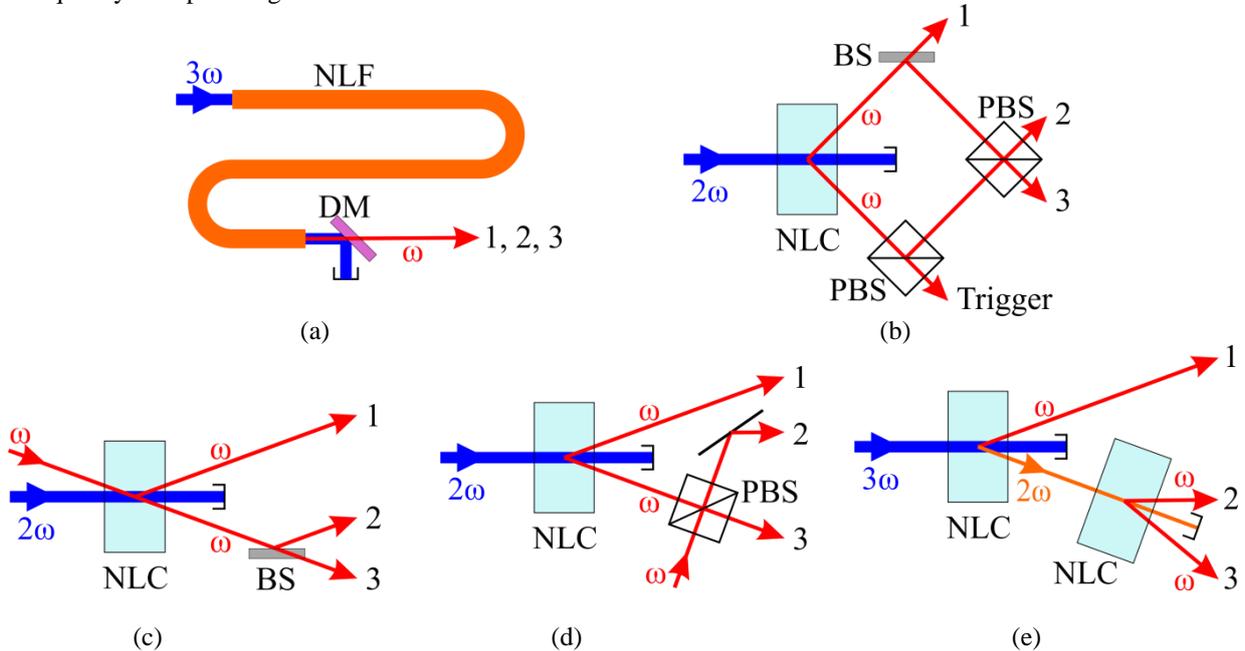

Figure 1. Triphoton generation schemes: (a) third order SPE in a waveguide; (b) the four-photon state selection; (c) and (d) interference scheme with a quasi-single photon source before and after SPE, respectively; (e) cascading generation.

## 3. POLARIZATION STATES OF TRIPHOTONS

There exist two classes among the pure polarization states of triphotons [25]: GHZ[26] and W[27] states.
States belonging to the first one are maximally entangled as described by the Bell inequalities, however if one is to generate a triplet in that state and detect one of the photons, then the state of the two others becomes separable
The states corresponding to the second class are not entangled, but after we detect one photon of the triplet in such state, then in 2/3 cases the others remain entangled. Both of those states have been realized experimentally [6,16–22]. While the measured fidelity lied in the range $F = 0.68 \div 0.86$, the interference visibility $V = \frac{I_{max} - I_{min}}{I_{max} + I_{min}}$ was in the range $V = 0.70 \div 0.86$ and the purity $P = 0.77 \div 0.88$ [26,27]. The total amount of registered photons varied from a few hundred to 30 thousand. It should also be noted that in all experiments triphotons were split into three different channels.

Thus, all three photons are distinguishable (so called nondegenerate case) and their quantum states are represented by a vector in an eight dimensional Hilbert space. At the same time, if the triphotons are generated due to TOSPDC, all triphotons may belong to one spatial and frequency mode (so called degenerate case). States such as $|HHV\rangle$, $|HVH\rangle$ и $|VHH\rangle$ are not distinguishable and the overall quantum state is described by a vector in a four dimensional space. Such states are described as triphoton polarization ququart[28].

## 4. QUANTUM TOMOGRAPHY PROTOCOLS AND THEIR PRECISION

According to Bohr's complementarity principle[29], different projection measurements form a set of mutually complementary measurements[30,31].
Thus the set of quantum measurements of a protocol forms a quantum measurement protocol and can be presented in the matrix form

$$M_j = X_{jl} c_l , \; j = 1, 2, ..., m, \tag{1}$$

where $c_l - l = 0, 2, ..., s-1$ are the components of a state vector in an $s$-dimensional Hilbert space, $M_j$ is the amplitude of the probability of the quantum projection with index $j$, $X_{jl}$ specifies the so-called instrumental matrix of a quantum measurement protocol. The protocol describes $m$ projections of a quantum state.

The parameters $\lambda_j$ specify the event generation intensities and $M_j$ is the amplitude:

$$\lambda_j = |M_j|^2. \tag{2}$$

The corresponding operator of intensity of the quantum process[2,3] is

$$\Lambda_j = X_j^+ X_j. \tag{3}$$

The event generation intensities can be then expressed as

$$\lambda_j = \text{Tr}(\Lambda_j \rho)$$

Our primary problem is to find the vector $c$ of the state which provides the maximum of the likelihood function.

In our case the likelihood function is defined by a product of Poisson probabilities $P(k_j) = \frac{(\lambda_j t_j)^{k_j}}{k_j!} \exp(-\lambda_j t_j)$ over all rows of protocol.

Here $k_j$ – the number of events obtained in the experiment, $t_j$ – the exposure time of row $j$.
With this notation the likelihood function is:

$$L = \prod_{j=1}^{m} \frac{(\lambda_j t_j)^{k_j}}{k_j!} \exp(-\lambda_j t_j). \tag{4}$$

A necessary condition for extremum of function leads to the likelihood equation[32].

$$Ic = Jc. \tag{5}$$

This sequence (1-6) allows for an efficient and a fast converging iterative procedure.
Here $I$ and $J$ are matrices with dimensions $s \times s$:

$$I = \sum_{j=1}^{m} t_j \Lambda_j, \quad J = \sum_{j=1}^{m} \frac{k_j}{\lambda_j} \Lambda_j. \tag{6}$$

That method of pure state reconstruction can be directly generalized to arbitrary mixed states.
The column vector $c$ of length $s$ is replaced by the matrix $c$ with size $s \cdot r$, where $r$ - is the evaluated number of components in the mixture that specifies the mixed state of the given rank, i.e. the number of nonzero eigenvalues of density matrix. The case $r = 1$ corresponds to the pure state whereas $r = s$ - to the mix of full rank. The matrix $c$ is a purified amplitude of states and its density matrix is $\rho = cc^+$.

The difference between the reconstructed and the true states can be attributed to the influence of statistical fluctuations related to the fundamental probabilistic nature of quantum mechanics.
The matrix of full probability in the form, given in[30,31] is a mathematical instrument for the quantitative description of fluctuation levels:

$$H = 2\sum_j \frac{t_j (\Lambda_j c)(\Lambda_j c)^+}{\lambda_j}. \tag{7}$$

Matrix $H$ is a real symmetric matrix which dimensions are $2rs \cdot 2rs$. It acts as a measure of information about parameters of a quantum state contained in the measurements determined by the tomography protocol. Above we have described complete tomographic protocols, i.e., protocols that provide arbitrarily accurate recovery of any mixed state as the sample size grows[28].

In case of a full protocol matrix $H$ has $\nu_H = (2s-r)r$ nonzero and strictly positive eigenvalues and the last $r^2$ eigenvalues are definitely equal to zero. On the basis of the information matrix we can show that the loss of accuracy is a random variable[32–35] and we can represent the asymptotic distribution as $1-F = \sum_{j=1}^{\nu} d_j \xi_j^2$, where $d_j \geq 0$ are nonzero coefficients, $\xi_j \sim N(0,1)$ and $j = 1,...,\nu$ are independent and normally distributed random variables with zero mean and unit variance. $v = v_H - 1 = (2s-r)r - 1$ is the number of degrees of freedom of the quantum state. For pure states $v = 2s - 2$ due to $r = 1$ and for mixed states of full rank $v = s^2 - 1$ and $r = s$.

The exact matching probability between reconstructed and theoretical states (Fidelity) is given by the following formula for pure states $F = |\langle c_{rec} | c_{theor} \rangle|^2$.

Similarly, for mixed states $F = Tr\left(\sqrt{\sqrt{\rho_{theor}} \rho_{rec} \sqrt{\rho_{theor}}}\right)^2$, where $\rho_{theor}$ is the theoretical density matrix and $\rho_{rec}$ is the reconstructed density matrix. The algorithm for calculating the coefficients $d_j$ $j = 1,...,v$ [28].

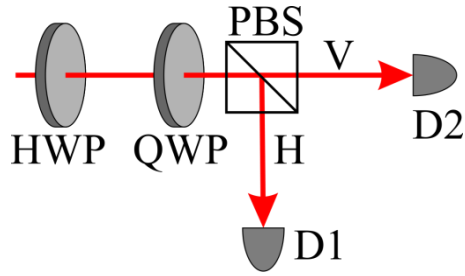

Figure 2. The experimental setup for one qubit tomography. HWP and QWP are half and quarter wave plates respectively, PBS is a polarization beam splitter, D1 and D2 are single-photon detectors.

## 5. NONDEGENERATE TRIPHOTON STATE TOMOGRAPHY

The typical experimental setup does not allow for projection measurements on entangled states, so we consider protocols with projections on the separable states. The most convenient way to create such complex protocols is to use products of one qubit tomography protocols.

For quantum tomography of two and more qubits it is common to use the tensor product of projection matrices. Different protocols (and so the matrices for evaluation) for each qubit can be used.

In every optical channel we place a pair of quarter (QWP) and half (HWP) wave plates realizing a given polarization transform, polarization beam splitter (PBS) and a pair of detectors (D1 and D2) as shown in Fig. 2.

Highly symmetrical solids can be used to construct one-qubit protocols[34]. Symmetrycal polyhedra are used to get the most uniform possible coverage of the Bloch sphere. States used for projective quantum measurements are determined by the direction out of the center of the Bloch sphere to the centers of the polyhedra faces .
Thus, the number of polyhedral faces determines the number of rows of one qubit protocol of quantum measurement.
Let us set the state $|V\rangle$ as the level of the logical zero and the state $|H\rangle$ as the logical unity, so $|0\rangle = |V\rangle$, $|1\rangle = |H\rangle$.
The multi qubit protocols of nondegenerate state tomography are formed by projective quantum measurements on corresponding states that are the tensor product of one qubit states under consideration. If one qubit protocol is based on a polyhedral with $m$ edges and so it has $m$ rows, then the corresponding $l$ - qubit protocol then has $m^l$ rows.

In this study, in order to simulate quantum tomography we have used three protocols of quantum measurements based on the geometry of the regular polyhedrons: tetrahedron, cube and octahedron protocols. These protocols are described by the following instrumental matrices:

$$X^{tetra} = \frac{1}{12^{1/4}} \begin{pmatrix} \sqrt{\sqrt{3}+1} & e^{i\pi/4}\sqrt{\sqrt{3}-1} \\ \sqrt{\sqrt{3}+1} & e^{i5\pi/4}\sqrt{\sqrt{3}-1} \\ \sqrt{\sqrt{3}-1} & e^{i3\pi/4}\sqrt{\sqrt{3}+1} \\ \sqrt{\sqrt{3}-1} & e^{i7\pi/4}\sqrt{\sqrt{3}+1} \end{pmatrix}, \; X^{cube} = \frac{1}{\sqrt{2}} \begin{pmatrix} \sqrt{2} & 0 \\ 0 & \sqrt{2} \\ 1 & 1 \\ 1 & -1 \\ 1 & i \\ 1 & -i \end{pmatrix}, \; X^{octa} = \frac{1}{12^{1/4}} \begin{pmatrix} \sqrt{\sqrt{3}+1} & e^{i\pi/4}\sqrt{\sqrt{3}-1} \\ \sqrt{\sqrt{3}+1} & e^{i3\pi/4}\sqrt{\sqrt{3}-1} \\ \sqrt{\sqrt{3}+1} & e^{i5\pi/4}\sqrt{\sqrt{3}-1} \\ \sqrt{\sqrt{3}+1} & e^{i7\pi/4}\sqrt{\sqrt{3}-1} \\ \sqrt{\sqrt{3}-1} & e^{i\pi/4}\sqrt{\sqrt{3}+1} \\ \sqrt{\sqrt{3}-1} & e^{i3\pi/4}\sqrt{\sqrt{3}+1} \\ \sqrt{\sqrt{3}-1} & e^{i5\pi/4}\sqrt{\sqrt{3}+1} \\ \sqrt{\sqrt{3}-1} & e^{i7\pi/4}\sqrt{\sqrt{3}+1} \end{pmatrix} \qquad (8)$$

## 6. THE RESULTS OF NUMERICAL EXPERIMENTS

For example let us reconstruct a three photon $GHZ$ state $|GHZ\rangle = \frac{1}{\sqrt{2}}(|HHH\rangle + |VVV\rangle)$ from a sample of size $n = 10^5$. In Fig. 3 we have plotted the values of fidelity loss distribution for protocols based on tetrahedron, cube and octahedron

The value of fidelity can lie in a wide interval, thus it is convenient to use a new variable $z = -\log_{10}(1-F)$. Here and below $\log_{10}$ denotes the common logarithm. The new variable $z$ defines the number of nines in numerical representation of fidelity, for example, $z = 4$ means that $F = 0.9999$.

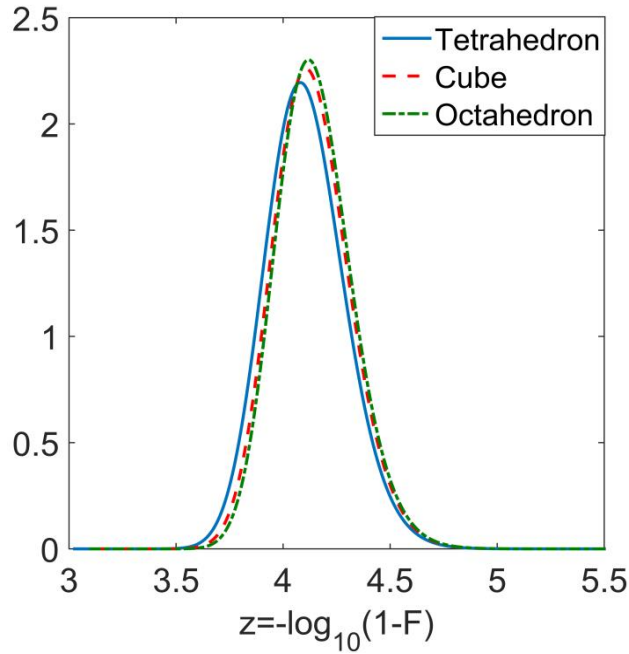

Figure 3. Probability density function of the number of nines in numerical representation of fidelity for various tomography protocols

As shown in Fig. 3 the reconstruction accuracy of quantum state is better for a protocol based on an octahedron. That means that the octahedron covers the Bloch sphere better than cube or tetrahedron. And if we use polyhedra with increasing amount of faces then the accuracy of restoration will increase. The average fidelity based on theory of precision for tetrahedron protocol is 99.991%, for octahedron protocol - 99.9922%. We use the fidelity loss function $L = n\langle 1-F \rangle$ that was introduced in[34] to characterize the asymptotic precision of tomography protocols.

The minimal achievable losses for pure states are:

$$L_{min} = s - 1. \qquad (9)$$

For three photon states $L_{min} = 7$. The numerical evaluation shows that the minimal achievable loss $L_{min}$ is determined by the theoretical limit (9). The upper bounds $L_{max}$ are the results of numerical optimization. The value of maximal losses $L_{max}$ and losses in $GHZ$ state $L_{GHZ}$ for tetrahedron protocol are 10.4 and 8.63 correspondly, while for octahedron protocol - 7.9 and 7.73 respectively.

Fig. 4a shows the distribution of accuracy for tetrahedron protocol with sample size $n = 10^5$, where the histogram is based on 200 numerical experiments and the curve is the theoretical distribution of fidelity loss. As can be seen, the experiments are in a good agreement with the theoretical distribution. The average fidelity $\langle F \rangle$, as indicated above, is 99.991%. If we restore the pure state as a maximally mixed one, i.e. it taking inadequate model, the average fidelity is 99.278%. Thus the fidelity loss accuracy increases by a factor of 80.

Let us now consider the reconstruction of mixture of a GHZ state and a state for which density matrix is proportional to the identity matrix:

$$\rho = f\frac{I}{s} + (1-f)|GHZ\rangle\langle GHZ|, \qquad (10)$$

where $I$ is a unit matrix with dimensions $8 \times 8$, $f$ is the weight of the maximally mixed state, $s$ is the dimension of the Hilbert space, for nondegenerate case $s = 8$ and for degenerate case $s = 4$ respectively. In our case (experiment) $f = 0.5$. The sample size is $n = 10^5$, 200 numerical experiments were performed with the tetrahedron protocol.

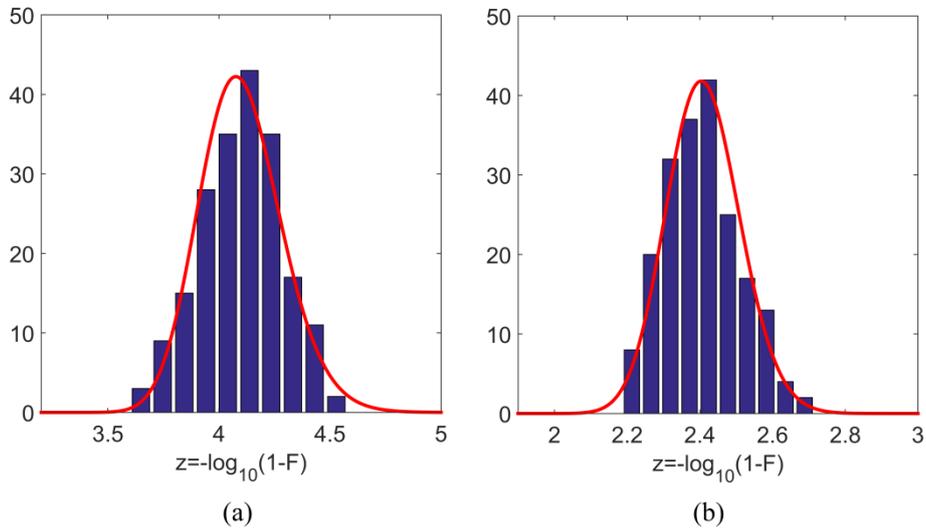

(a)  (b)

Figure 4. Probability density function of the number of nines in numerical representation of fidelity for the (a) pure state and for the (b) mixed state(10).

The average fidelity for numerical results $\langle F \rangle$ is 99.596% . The lower values of fidelity are explained by the need to measure $s^2 - 1 = 63$ parameters of the mixed state, as opposed to $2s - 2 = 14$ parameters of a pure state.

## 7. DEGENERATE TRIPHOTON STATE TOMOGRAPHY

In the case of degenerate three photon state tomography (see Fig. 2) quantum states are invariant with respect to photon permutations. To describe that let us select the reduced basis of four vectors:

$$
\begin{aligned}
|\psi_0\rangle &= |3_V\rangle = |000\rangle = |VVV\rangle \\
|\psi_1\rangle &= |2_V, 1_H\rangle = \frac{1}{\sqrt{3}}(|001\rangle + |010\rangle + |100\rangle) = \frac{1}{\sqrt{3}}(|VVH\rangle + |VHV\rangle + |HVV\rangle) \\
|\psi_2\rangle &= |1_V, 2_H\rangle = \frac{1}{\sqrt{3}}(|011\rangle + |101\rangle + |110\rangle) = \frac{1}{\sqrt{3}}(|VHH\rangle + |HVH\rangle + |HHV\rangle) \\
|\psi_3\rangle &= |3_H\rangle = |111\rangle = |HHH\rangle
\end{aligned}
\quad (11)
$$

All possible superpositions of states (9) span a state space of a three-photon polarization ququart[28]. Vector columns of previously described states form the matrix $G$ that performs the transition from the four-dimensional basis to original eight-dimensional one.

$$c_{4d} = G^\dagger c_{8d} \quad (12)$$

Here $c_{8d}$ is a column vector of 8 elements and $c_{4d}$ is a column vector of 4 elements. The 8-by-4 matrix $G$:

$$
G = \begin{pmatrix}
1 & 0 & 0 & 0 \\
0 & 1/\sqrt{3} & 0 & 0 \\
0 & 1/\sqrt{3} & 0 & 0 \\
0 & 0 & 1/\sqrt{3} & 0 \\
0 & 1/\sqrt{3} & 0 & 0 \\
0 & 0 & 1/\sqrt{3} & 0 \\
0 & 0 & 1/\sqrt{3} & 0 \\
0 & 0 & 0 & 1
\end{pmatrix}
\quad (13)
$$

This matrix $G$ allows us to transform the original m-by-8 instrumental matrix $X_{8d}$ to reduced m-by-4 matrix $X_{4d}$

$$X_{4d} = X_{8d} G \quad (14)$$

In the new basis the equation (1) becomes:

$$M_j^{4d} = X_{jl}^{4d} c_l^{4d}, \quad j = 1, 2, \ldots, m, \quad (15)$$

The solution to the initial state reconstruction in a reduced basis is similar to the one in the original basis, but lets us take into account the smaller number of parameters. In Fig.5a the distribution of the number of nines in numerical representation of fidelity for four dimensional state is shown. The sample size is $n=10^5$, the line shows the theoretical distribution of fidelity losses and the histogram shows the result of 200 numerical experiments (evaluations). We used the tetrahedron protocol. The average fidelity is 99.996%. If we restore the pure state as an absolutely mixed one the average fidelity is 99.669%. Thus the fidelity losses increase by a factor of 83.

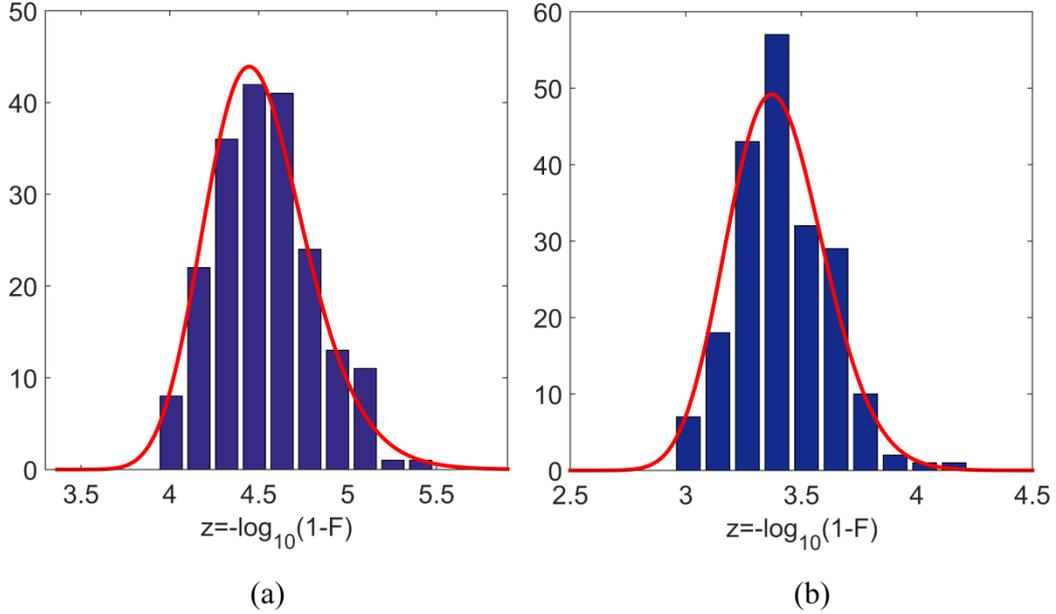

Figure 5. The probability distribution of the number of nines in numerical representation of fidelity for (a) pure GHZ state reconstruction in four dimensions and for (b) the mixed state(10) in four dimensions.

The average fidelity is 99.956%. In the transition from an eight-dimensional space into a four-dimensional space the accuracy losses are reduced 9 times. In both cases the reconstruction accuracy reaches 99.99%. The corresponding results of previously mentioned publications are 86% and 82%[26,27]. Thus we can conclude that the significant errors were made in generation and measurement of GHZ states.

## CONCLUSION

We analyze different tomography protocols for polarization triphoton GHZ states that provide precision close to the theoretically possible maximum. We show in a numerical experiment of significant increase in tomography accuracy in case of degenerate triphoton basis for a pure GHZ state. We use our findings to develop a fundamental and experimental approach to analyze quantum protocols that we have used previously for reconstruction of GHZ state.

The work was supported by the Russian Science Foundation, grant no 14-12-01338.